\pdfoutput=1
\documentclass{iopart}
\usepackage{graphicx,amssymb}
\usepackage{tikz}
\usepackage[english]{babel}

\newcommand{\diff}{\mathrm{d}}
\newcommand{\eqref}[1]{(\ref{#1})}

\begin{document}
\title{A generalisation of Schramm's formula for SLE$_2$}
\author{Christian Hagendorf}%
\address{CNRS-Laboratoire de Physique Th\'eorique, Ecole Normale
Sup\'erieure, 75231 Paris cedex 05, FRANCE}
\ead{hagendor@lpt.ens.fr}

\date{\today}

\begin{abstract}
 The scaling limit of planar loop-erased random walks is described by a stochastic Loewner evolution with parameter $\kappa=2$. In this note SLE$_2$ in the upper half-plane $\mathbb{H}$ minus a simply-connected compact subset $\mathbb{K}\subset \mathbb{H}$ is studied. As a main result, the left-passage probability with respect to $\mathbb{K}$ is explicitly determined.
 \end{abstract}

\maketitle

\section{Introduction}

The study of the scaling limit of loop-erased random walks (LERWs) has incited the invention of stochastic Loewner evolutions (SLEs) by Schramm \cite{schramm:00}. In fact, SLE$_\kappa$ is a one-parameter family of conformally invariant measures on non-self-crossing planar curves. Lawler, Schramm and Werner showed that the scaling limit of LERWs in simply-connected planar domains is SLE with $\kappa=2$ \cite{lsw:04}. The corresponding problem in multiply-connected domains has been investigated in a series of works by Dapeng Zhan \cite{zhan:04,zhan:06,zhan:06_2}. Besides LERWs, interfaces in a variety of models in statistical mechanics at criticality can be identified as SLEs. For example, domain walls in the critical two-dimensional Ising model correspond to $\kappa=3$, critical percolation hulls are described by $\kappa=6$ and uniform spanning trees by $\kappa=8$.

To be more precise, the (chordal) SLE measure $\mu_{\mathbb{D},x_0,x_\infty}$ on random curves (or hulls) $\gamma$ in a planar domain $\mathbb{D}$ with endpoints $x_0,\,x_\infty\in \partial\mathbb{D}$ on the boundary is characterised by the following two properties. \textit{(i) Conformal invariance:} If $f$ is a conformal map from $\mathbb{D}$ to $\mathbb{D}'$ then the image of SLE in $\mathbb{D}$ is SLE in $\mathbb{D}'$: $f\circ \mu_{\mathbb{D},x_0,x_\infty} = \mu_{\mathbb{D}',f(x_0),f(x_\infty)}$. Conformal invariance therefore always allows to study SLE in a suitable reference domain. \textit{(ii) Domain Markov property:} if we condition on a first portion of the curve, say $\gamma'$ from $x_0$ to $z\in \mathbb{D}$, then the remainder $\gamma\backslash\gamma'$ has the law $\mu_{\mathbb{D}\backslash\gamma',z,x_\infty}$. The curves/hulls may be studied by Loewner's method of slit mappings \cite{ahlfors:73}. For simply-connected domains, if we make the standard choice $\mathbb{D}=\mathbb{H}$ and $x_0=0$, $x_\infty=\infty$, and parametrise the curves $\gamma_t$ by a real parameter $t\geq 0$, then the conformal mapping $g_t:\mathbb{H}\backslash \gamma_t \to \mathbb{H}$ is solution of the \textit{Loewner differential equation}
\begin{equation}
  \label{eqn:loewnerhp}
  \frac{\diff g_t(z)}{dt}=\frac{2}{g_t(z)-W_t}, \qquad g_0(z)=z.
\end{equation}
Here, the so-called \textit{driving process} $W_t\in\mathbb{R}$ is the image of the tip of $\gamma_t$ (stricto sensu this form of Loewner's equation is valid only for the particular parametrisation by half-plane capacity, see e.g. the book \cite{lawler:05} for a detailed description).
The conditions \textit{(i)} and \textit{(ii)} imply that $W_t=\sqrt{\kappa}B_t+\alpha t$ where $B_t$ a is standard Brownian motion, and $\kappa>0$ and $\alpha$ real parameters. Reflection symmetry, as it occurs usually in statistical mechanics, leads to $\alpha=0$.

SLE not only provides a sound mathematical basis for the analysis of problems formerly addressed by boundary conformal field theory (boundary CFT), but also sheds new light on CFT-genuine questions. Whereas CFT mainly focusses on correlation functions of \textit{local} operators (such as spins), SLE describes extended geometric objects like interfaces and hulls. The understanding of the interplay of these two aspects of conformally invariant models in statistical mechanics has led to the so-called SLE/CFT correspondence according to which CFT correlation functions correspond to SLE martingales and probabilities \cite{bauer:03,bauer:06}. A paradigm for the latter is Schramm's formula \cite{schramm:01}: if we consider chordal SLE$_\kappa$ in $\mathbb{H}$ and a point $z\in \mathbb{H}$ then the trace may pass either to the left or to the right of $z$. Let us write $\mathcal{L}_z$ for the event of left-passage. Then it is possible to use \eqref{eqn:loewnerhp} and the SLE properties in order to show
\begin{equation}
  \label{eqn:schramm}
  \mathsf{P}[\mathcal{L}_z] = \frac{1}{2}+\frac{\Gamma(4/\kappa)}{\sqrt{\pi}\,\Gamma((8-\kappa)/2\kappa)}\frac{x}{y}\,{}_2F_1\left(\frac{1}{2},\frac{4}{\kappa},\frac{3}{2};-\frac{x^2}{y^2}\right).
\end{equation}
${}_2F_1(a,b,c;z)$ denotes the hypergeometric function \cite{abramowitz:70}. Indeed it can be shown that this result coincides with a correlation function of boundary CFT \cite{bauer:03}.

\begin{figure}[h]
\centering
\begin{tabular}{cp{0.5cm}c}
\begin{tikzpicture}[>= stealth,scale=0.8]
 \filldraw[lightgray] (0,0) rectangle (6,-0.2);
 \draw[thick] (0,0)--(6,0);
 \draw[fill=lightgray,rounded corners,thick] (3,1.5) .. controls (3.4,1.3) and (3.5,1.7) .. (3.6,2) .. controls (3.2,2.7) and (3.5,3) .. (2.3,2.8) .. controls (1.7,2.5) and (2,2) .. (3,1.5);
  \draw[rounded corners, red, thick] (3,0) .. controls (3,1) and (2.5,1) .. (1.5,1.5) .. controls (1,2) and (1.8,2) .. (1.7,2.7) ..  controls (1.7,3.5) and (1.7,3.5) .. (2.3,4);
  \draw[rounded corners, red, thick, dashed] (3,0) .. controls (3.1,1.5) and (3.5,0.6) .. (4.5,1) .. controls (5,2) and (5,2) .. (4,2.5) .. controls (3.7,3.5) and (3.3,3.5) .. (3.5,4);
 \draw (3,-0.4) node{$0$};
 \draw (2.7,2.3) node{$\mathbb{K}$};
 \draw (0.5,3.5) node{$\mathbb{H}$};
 \filldraw (3,0) circle (1.5pt);
\end{tikzpicture}
&&
\begin{tikzpicture}[>= stealth]
  \filldraw[color=lightgray](0,0) rectangle (5,-0.1);
  \filldraw[color=lightgray](0,2.5) rectangle (5,2.6);
  \draw[thick](0,0) -- (5,0);
  \draw[thick](0,2.5) -- (5,2.5);
  \draw(2.5,-0.4) node{$0$};
  \draw(0,-0.4) node{$-\pi$};
  \draw(5,-0.4) node{$\pi$};
  \draw(4,-0.4) node{$x$}; 
  \draw[rounded corners,red, dashed, thick] (2.5,0) .. controls (2.9,1) and (2.5,1.5) .. (3,1.5) .. controls (4,0.5) and (3,0.25).. (4,0);
  \draw[rounded corners,red, thick] (2.5,0) .. controls (2.,1.5) and (2.5,2.3) .. (1.5,2) .. controls (0.85,1.05) and (1,1.5) ..(0,1.5);
  \draw[rounded corners,red, thick] (5,1.5) .. controls (4.75,1.5) and (4.75,1.5) .. (4.25,2) .. controls (3,1.5) and (4.3,1) .. (4,0);
  \draw[<->](-0.25,0) -- (-0.25,2.5); \draw(-0.5,1.25) node{$p$};
  \draw[densely dotted,thick] (0,0) -- (0,2.5);
  \draw[densely dotted,thick] (5,0) -- (5,2.5);
  \draw (0,1.25) node[rotate=40] {$=$};
  \draw (5,1.25) node[rotate=40] {$=$}; 
  \filldraw (2.5,0) circle (1.5pt);
  \filldraw (4.,0) circle (1.5pt);
  \draw (0.5,0.4) node{$\mathbb{T}_p$};
\end{tikzpicture}\\
(a) && (b)
\end{tabular}
\caption{(a) Two ways going around a simply-connected compact subset $\mathbb{K}\subset \mathbb{H}$: the trace may pass to the left (solid line) or to the right (dashed line) (b) The analogous situation on the cylinder $\mathbb{T}_p$.}
\label{fig:annulus}
\end{figure}
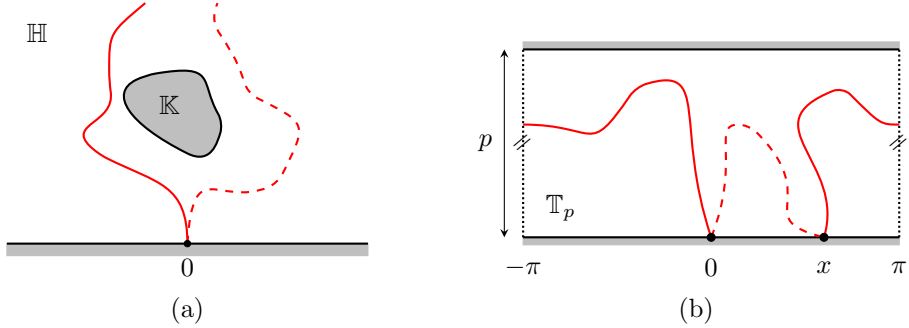

In this work we consider the following generalisation of \eqref{eqn:schramm}. Consider a simply-connected compact subset $\mathbb{K}\subset \mathbb{H}$ as shown on figure \ref{fig:annulus}a. Let us study SLE$_\kappa$ traces in $\mathbb{D}=\mathbb{H} \backslash \mathbb{K}$ from $x_0=0$ to $x_\infty=\infty$ what amounts to a version of chordal SLE in a \textit{doubly-connected domain} \cite{rbauer:08}.
Denote $\mathcal{L}_\mathbb{K}$ the event that the traces go to the left with respect to $\mathbb{K}$ and compute the probability $\mathsf{P}[\mathcal{L}_\mathbb{K}]$. As we shrink $\mathbb{K}$ to a point we expect to recover \eqref{eqn:schramm}. The main purpose of the following considerations is to work out this passage probability in the case of LERWs where $\kappa=2$ which turns out to be amenable to explicit computations.

The article is organised as follows. In section \ref{sec:results} we present a brief sketch of SLE$_2$ in doubly-connected domains, provide Loewner's equation, and state the main results of this work. In section \ref{sec:ward} we use the SLE/CFT correspondence in order to work out the drift of Loewner's equation from the conformal Ward identities. Section \ref{sec:lerw} is devoted to the application to the case $\kappa=2$. In particular, we provide the solution of the differential equation for the left-passage probability that is inspired by an analogy of diffusion-advection in a Burgers flow. Next, we consider an extension to side arcs as targets for the SLE$_2$ traces by employing the method of conditioning SLEs. After concluding in section \ref{sec:conclusion}, we recall the derivation of the excursion Poisson kernel, conformal Ward identities on doubly-connected domains and some considerations on the spectral determinant of the Laplacian related to our work in an appendix.

\section{Results}
\label{sec:results}

\paragraph{SLE on doubly-connected domains.} Conformal invariance always allows to study SLE in a reference geometry. Other geometries may then considered by conformal mapping. The domain $\mathbb{D}=\mathbb{H}\backslash \mathbb{K}$ has the topology of an annulus (its complement consists of two connected components, each of which having at least two boundary points). It is known that such a domain may be mapped onto an annulus $\mathbb{A}_p=\{z\in \mathbb{C}\mathop{|} e^{-p}<|z|<1\}$ where $p>0$ is some real number called \textit{modulus} of $\mathbb{D}$ (see for example \cite{lawler:05}). Two annuli $\mathbb{A}_p$ and $\mathbb{A}_q$ are conformally equivalent if and only if $p=q$ \cite{nehari:82}.
Although annuli $\mathbb{A}_p$ constitute a canonical choice as reference domains, we shall work out the problem rather for a cylinder geometry $\mathbb{T}_p$ as depicted on figure \ref{fig:annulus}b. We obtain the cylinder from an infinite strip $\mathbb{S}_p = \{z\in \mathbb{C}\mathop{|} 0 < \mathrm{Im}\, z < p\}$ by identifying any two $z,w\in\mathbb{S}_p$ if $\mathrm{Re}\,(z-w) \in 2\pi \mathbb{Z}$. We denote by $\partial_1 \mathbb{T}_p$ the ``lower'' boundary, and by $\partial_2 \mathbb{T}_p$ its ``upper'' boundary. Hence for any $\mathbb{K}$ there is a conformal mapping $f:\mathbb{H}\backslash \mathbb{K}\to \mathbb{T}_p$ with some suitable $p$ such that $f(\mathbb{R})=\partial_1\mathbb{T}_p$ and $f(\partial\mathbb{K})=\partial_2\mathbb{T}_p$. Because of translation invariance along the real axis, we may choose $f$ such that $f(0)=0$. Moreover we shall denote the image of $\infty$ by $x$. For example, if $\mathbb{K}=\bar{\mathbb{B}}_r(z_0)=\{z\in \mathbb{H}\mathop{|}|z-z_0|\leq r\}$ is a closed disk centred at $z_0=x_0+iy_0,\, y_0>0$, we have $\cosh 2p=(y_0/r)^2$ and $\cot x/2 = -x_0/\sqrt{y_0^2-r^2}$.

\begin{figure}[h]
  \begin{center}
  \begin{tabular}{cp{1cm}c}
    \includegraphics[width=0.4\textwidth]{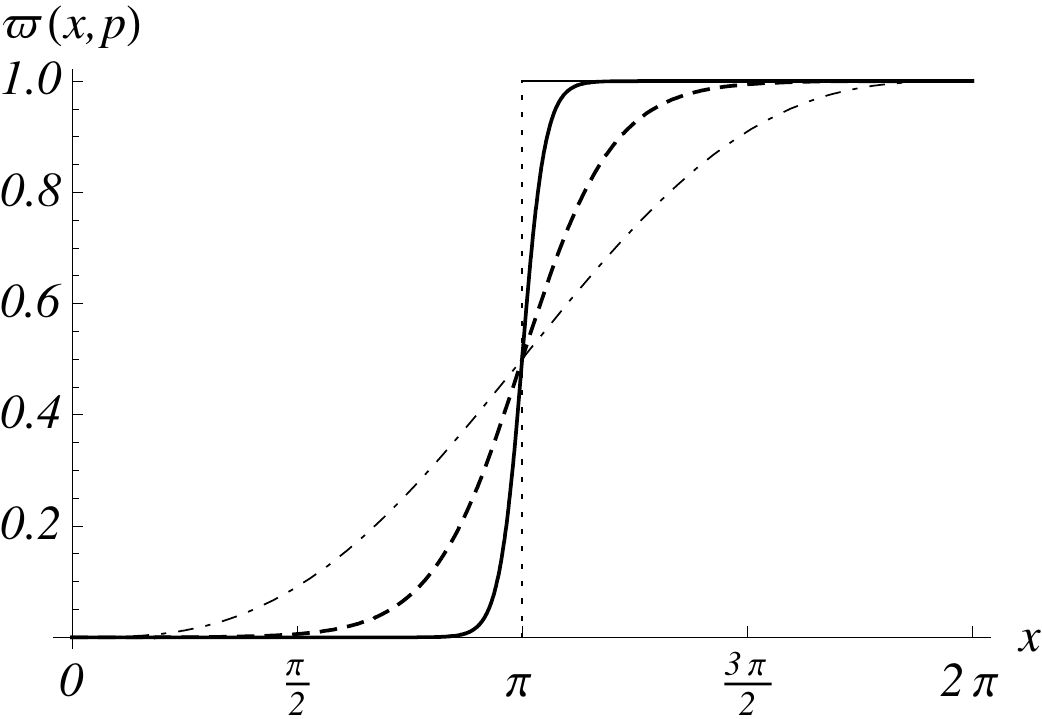}&&
    \includegraphics[width=0.4\textwidth]{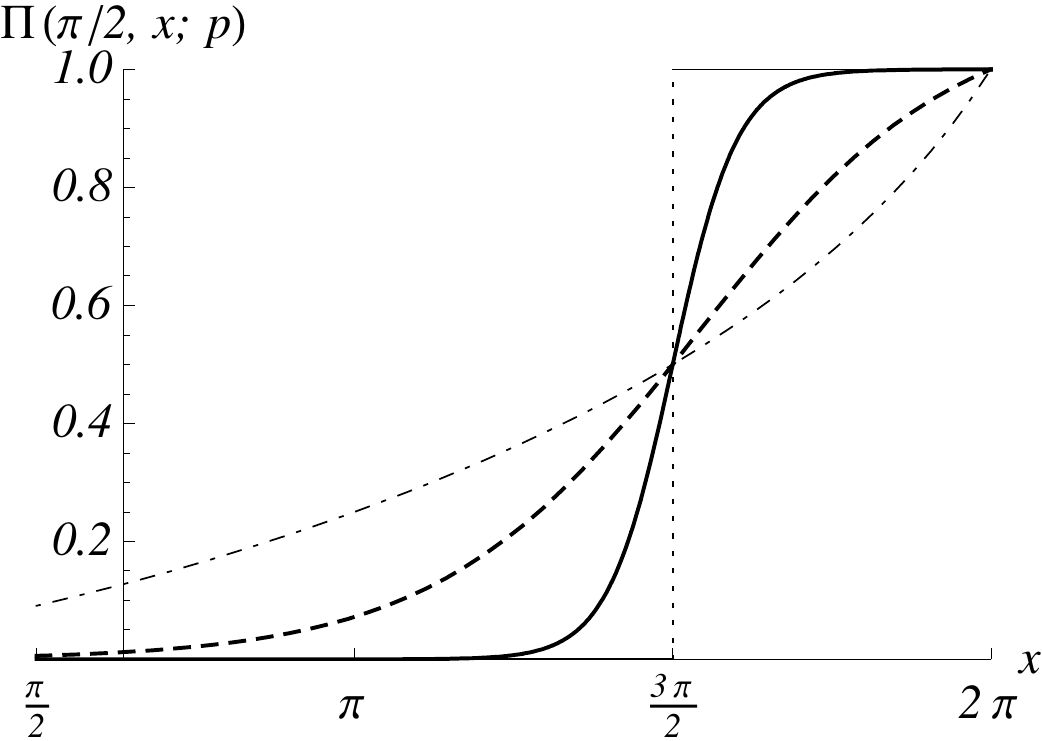}\\
    (a) & &(b)
  \end{tabular}
  \end{center}
  \caption{Illustration of the probabilities (a) $\varpi(x,p)$ and (b) $\Pi(p;\pi/2,x)$ for $p=1/2$ (solid) and $p=2$ (dashed) together with the asymptotic cases for $p\to 0^+$ (unit step) and $p\to \infty$ (dot-dashed).}
  \label{fig:proba}
\end{figure}

We shall therefore study SLE on $\mathbb{T}_p$. \textit{Loewner's differential equation} (and hence the equivalent to \eqref{eqn:loewnerhp}) is given by \cite{zhan:06, bauer:04_2}
\begin{equation}
  \label{eqn:loewner}
  \frac{\diff g_t(z)}{\diff t} = v(g_t(z)-W_t, p-t),\qquad g_{t=0}(z)=z,\, z\in \overline{\mathbb{T}}_p.
\end{equation}
Here, the time parametrisation is chosen in such a way that $g_t(z)$ maps $\mathbb{T}_p\backslash \gamma_t$ to $\mathbb{T}_{p-t}$, $0\leq t < p$. The vector field $v(z,p)$ may be written as the logarithmic derivative of the Jacobi theta function $\theta_1(z|\tau)=-i\sum_{n\in \mathbb{Z}}(-1)^n y^n q^{(n+1/2)^2/2}$ with $y=e^{2\pi iz}$, $q=e^{2\pi i\tau}$ (the prime denotes the derivative with respect to the first argument):
\begin{equation}
  \label{eqn:defv}
   v(z,p) = \frac{1}{\pi}\frac{\theta_1'(z/2\pi|ip/\pi)}{\theta_1(z/2\pi|ip/\pi)} = \cot\left(\frac{z}{2}\right)+4\sum_{n=1}^\infty\frac{\sin nz}{e^{2np}-1},
\end{equation}
Zhan has considered SLE traces from $z=0$ to $\partial_2\mathbb{T}_p$, \cite{zhan:06}. In this case $W_t =\sqrt{\kappa}B_t$ where $B_t$ is standard Brownian motion. However, in our problem the traces are suppose to go from $0$ to $x$. Therefore presence of a marked point $x$ on the boundary will lead to an additional drift of the driving process. In fact for $\kappa=2$, as we shall show in section \ref{sec:ward}, it is solution of the stochastic differential equation
\begin{equation}
  \label{eqn:driftsde}
  \diff W_t = \sqrt{2}\,\diff B_t-2\frac{\partial}{\partial y}\left[ \ln H(g_t(x)-W_t,p-t)\right]\,\diff t,\qquad W_0=0,
\end{equation}
where $H(y,p)$ denotes the \textit{excursion Poisson kernel} for $\mathbb{T}_p$ on $\partial_1\mathbb{T}_p$ (see \ref{app:green}). It is related to $v(y,p)$ by
\begin{equation}
  H(y,p) = -\frac{1}{\pi}\left(v'(y,p)+\frac{1}{p}\right).
\end{equation}
\paragraph{Passage probabilities.} After conformal mapping to $\mathbb{T}_p$ the left-passage with respect to $\mathbb{K}$ translates to a left-passage as illustrated on figure \ref{fig:annulus}b. We shall denote the corresponding event by $\mathcal{L}$. Using the preceding equations, we are going to show that the left-passage probability $\varpi(x,p)=\mathsf{P}_{x,p}[\mathcal{L}]$ on $\mathbb{T}_p$ for $\kappa=2$ is given in terms of the function $v(z,p)$ as
\begin{equation}
  \varpi(x,p) = \frac{1}{2\pi}\left(x+\frac{\partial(p\, v(x,p))/\partial p}{\,v'(x,p)+1/p}\right).
\end{equation}
$\varpi(x,p)$ for different moduli $p$ is illustrated on figure \ref{fig:proba}. This formula can be extended to the case if the target is a side arc $[a,b]\in \partial_1\mathbb{T}_p$. In this case, we are going to show that the left-passage probability $\Pi(p;a,b)$ is given by
\begin{equation}
 \Pi(a,b;p) =\frac{\Omega(b,p)-\Omega(a,p)}{p(v(b,p)-v(a,p))+b-a},
\end{equation}
where $\Omega(x,p)$ may be written in terms of $v(x,p)$ as
\begin{equation}
\fl\Omega(x,p)=\frac{(x-\pi)(x+\pi)+2p}{4\pi}+\frac{p\,x}{2\pi}\,v(x,p)+\frac{p^2}{2\pi}\left(v'(x,p)+\frac{1}{2}v(x,p)^2\right).
\label{eqn:defomega}
\end{equation}
An illustration for $\Pi(\pi/2,x;p)$ for different moduli is shown on figure \ref{fig:proba}b.

\section{SLE/CFT correspondence and boundary correlation functions}
\label{sec:ward}
In this section we shall justify the stochastic differential equation \eqref{eqn:driftsde} for the driving process $W_t$. We shall use an argument from the SLE/CFT correspondence, namely the so-called statistical martingale trick (see e.g. the review \cite{bauer:06} for a detailed explanation).

\subsection{SLE/CFT correspondence}
According to the SLE/CFT correspondence, SLE$_{\kappa}$ defined on a planar domain $\mathbb{D}$ describes interfaces in boundary CFTs with central charge $c=(6-\kappa)(3\kappa-8)/2\kappa$. SLE traces/hulls emerging from a boundary point $x \in \partial\mathbb{D}$ are implemented by boundary condition changing operators $\psi_{1,2}(x)$ with conformal weight $h_{1,2}(\kappa)=(6-\kappa)/2\kappa$. In fact, $\psi_{1,2}(x)$ is a Virasoro primary operator degenerate at level two \cite{bauer:03}. 

One of the main interests for characterising SLE traces consists in computing expectation values of observables $\mathcal{O}$, such as in our case the projector on the subset of traces that go to the left on $\mathbb{T}_p$. These expectation values are related to CFT correlation functions via
\begin{equation}
  \label{eqn:observable}
  \prec \mathcal{O}\succ_{\mathbb{T}_p,b.c.} = \frac{\langle \mathcal{O}\, \psi_{1,2}(0)\psi_{1,2}(x)\rangle_{\mathbb{T}_p}}{\langle \psi_{1,2}(0)\psi_{1,2}(x)\rangle_{\mathbb{T}_p}}.
\end{equation}
Here the subscript ``b.c.'' abbreviates ``boundary conditions''. In our case they correspond to the existence of a single interface with endpoints $0$ and $x$.
Let us now condition an a first portion $\gamma_t$ of the curve. If we wish to compute the expectation value of $\mathcal{O}$ in the cut domain $\mathbb{D}_t=\mathbb{T}_p\backslash \gamma_t$, we may use the Markov property and conformal invariance in order to map this problem back to a standard domain via the Loewner mapping $g_t(z)$. For simply-connected domains, the Riemann mapping theorem implies that we may choose $\mathbb{D}$. However, for doubly-connected domains such a mapping usually involves a change of the modulus \cite{rbauer:08}. Using the solutions of \eqref{eqn:loewner} we therefore have
\begin{equation}
  \prec \mathcal{O}\succ_{\mathbb{D}_t,b.c.} = \frac{\langle {}^{g_t}\mathcal{O}\, \psi_{1,2}(W_t)\psi_{1,2}(X_t) \rangle_{\mathbb{T}_{p-t}}}{\langle \psi_{1,2}(W_t)\psi_{1,2}(X_t)\rangle_{\mathbb{T}_{p-t}}},\qquad X_t = g_t(x).
  \label{eqn:martingale}
\end{equation}
Here $ {}^{g_t}\mathcal{O}$ is the image of $\mathcal{O}$ under $g_t(z)$. Notice in particular that left-passage is invariant under such a conformal transport. Now the SLE/CFT correspondence states that $\prec \mathcal{O}\succ_{\mathbb{D}_t,b.c.}$ has to be a local martingale for SLE. This is the statistical martingale trick. Therefore the drift term of its It\^o derivative must vanish. This is intimately related to the fact, that the level two degeneracy of $\psi_{1,2}$ leads to particular partial differential equations satisfied by the correlation functions in \eqref{eqn:observable}.

\subsection{Conformal Ward identities and differential equations}

A general strategy to find differential equations for CFT correlation functions is to first write the conformal Ward identities. They involve the stress tensor $T(z)$ and encode their transformation behaviour with respect to conformal mappings \cite{difrancesco:97}. To this end, let us consider the product $\mathcal{O}=\prod_{k=1}^n \varphi_{k}(z_k,\bar{z}_k)$ where $\varphi_k(z_k,\bar{z}_k)$ are spinless primary fields with conformal weights $h_k$ on $\mathbb{T}_p$. Since we work with a boundary conformal field theory we shall identify $\bar{z}_k$ with the complex conjugate $z_k^\ast$ of $z_k$ \cite{cardy:84}. Then it holds for an insertion of the stress tensor $T(z)$ at some point $z\in \mathbb{T}_p$:
\begin{eqnarray}
  \label{eqn:ward}
  \fl \langle T(z) \mathcal{O}\rangle_{\mathbb{T}_p}  =&\sum_{k=1}^n\left(-\frac{h_k}{2}v'(z-z_k,p)+\frac{v(z-z_k,p)}{2}\frac{\partial}{\partial  z_k}+(z_k\to \bar{z}_k)\right)\langle \mathcal{O}\rangle_{\mathbb{T}_p}\nonumber\\
 \fl  &+\frac{1}{2Z}\frac{\partial}{\partial p}\left(Z\langle\mathcal{O}\rangle_{\mathbb{T}_p}\right)
\end{eqnarray}
with the function $v(z,p)$ given in \eqref{eqn:defv}. $Z=Z(p)$ is the CFT partition function that depends on the modular parameter $p$ and on boundary conditions. In fact, \eqref{eqn:ward} corresponds the torus Ward identity found by Eguchi and Ooguri \cite{eguchi:87} for $2n$ points $(z_1,\dots,z_n,\bar{z}_1=z_1^\ast,\dots,\bar{z}_n=z_n^\ast)$. We shall give an elementary derivation of \eqref{eqn:ward} in \ref{app:ward}.

Using \eqref{eqn:ward} it is straightforward to derive a differential equation involving a primary field which is degenerate at level two such as
 $\psi_{1,2}(x)$ in \eqref{eqn:observable}. For simplicity, we shall only consider a correlation function involving a product of boundary Virasoro fields $\mathcal O=\prod_{k=1}^n \varphi_{k}(x_k),\, x_k\in \partial_1\mathbb{T}_p$. Then we find for the correlation function $\langle \psi_{1,2}(w)\psi_{1,2}(x)\mathcal{O}\rangle_{\mathbb{T}_p}$
\begin{eqnarray}
 \fl\left(\frac{\kappa}{2}\frac{\partial^2}{\partial w^2} +v(p,x-w)\frac{\partial}{\partial x}+h_{1,2}\left(v'(p,x-w)-\Xi(p)\right)\right)\langle\psi_{1,2}(w)\psi_{1,2}(x)\mathcal{O}\rangle_{\mathbb{T}_p}\nonumber\\ 
  = -\sum_{k=1}^n\left(h_k v'(p,x_k-w)+v(p,x_k-w)\frac{\partial}{\partial x_k}\right)\langle\psi_{1,2}(w)\psi_{1,2}(x)\,\mathcal O\rangle_{\mathbb{T}_p}\nonumber\\
 \qquad+\frac{1}{Z}\frac{\partial}{\partial p}\left(Z\langle\psi_{1,2}(w)\psi_{1,2}(x)\,\mathcal O\rangle_{\mathbb{T}_p}\right)
   \label{eqn:bdpde}
\end{eqnarray}
where we have abbreviated $\Xi(p) = 1/6-\sum_{n=1}^\infty 4n/(e^{2np}-1)$.

Let us use this equation in order to analyse the driving process $W_t$ for Loewner's equation \eqref{eqn:loewner}. We shall make the \textit{ansatz} $\diff W_t = \sqrt{\kappa}\,\diff B_t+ f(W_t, X_t)\,\diff t$ where $X_t=g_t(x)$. $f$ is a function to be determined from the requirement that \eqref{eqn:martingale} is a local martingale. If $\mathcal{O}$ again is a product of boundary primary operators it transforms according to
\begin{equation}
  {}^{g_t}\mathcal{O} = \mathcal{J} \prod_{k=1}^n\varphi_k(g_t(x_k)),\qquad \mathrm{with}\quad \mathcal{J} =\prod_{k=1}^n g_t(x_k)^{h_k}.
\end{equation}
In order to compute the derivatve we use Loewner's equation $\diff g_t(z) = v(g_t(z)-W_t,p-t)\,\diff t$ and $\diff g'_t(z)= g_t'(z) v(g_t(z)-W_t, p-t)\,\diff t$. Abbreviating $X_k = g_t(x_k)$ we find
\begin{equation}
  \fl\diff \left({}^{g_t}\mathcal{O}\right) = \mathcal{J}\sum_{k=1}^n\left(h_k v'(X_k-W_t,p-t)+v(X_k-W_t,p-t)\frac{\partial}{\partial X_k}\right)\prod_{k=1}^n\varphi_k(X_k).
\end{equation}
Hence, upon computing the It\^o derivative we find
\begin{eqnarray}
 \fl\diff \prec \mathcal{O}\succ_{\mathbb{D}_t,b.c.}
 = \mathcal{J}\Biggl(\sqrt{\kappa}\,\diff B_t\,\frac{\partial}{\partial W}\nonumber\\
 +\diff t\biggl[-\frac{\partial}{\partial p}+\frac{\kappa}{2}\frac{\partial^2}{\partial W^2}+f(W_t,X_t)\frac{\partial}{\partial W}+v(X_t-W_t,p-t)\frac{\partial}{\partial X}\\
 \fl+\sum_{k=1}^n\left(h_k v'(X_k-W_t,p-t)+v(X_k-W_t,p-t)\frac{\partial}{\partial X_k}\right)\biggr]\Biggr)\frac{\langle \mathcal{O}\, \psi_{1,2}(W_t)\psi_{1,2}(X_t) \rangle_{\mathbb{T}_{p-t}}}{\langle \psi_{1,2}(W_t)\psi_{1,2}(X_t)\rangle_{\mathbb{T}_{p-t}}}\nonumber
\end{eqnarray}
Since $\prec \mathcal{O}\succ_{\mathbb{D}_t,b.c.}$ has to be a martingale, the drift term must vanish. It turns out (after a little algebra) that this is only compatible with the null-vector equations \eqref{eqn:bdpde} if
\begin{equation}
f(W_t,X_t) = -\kappa\frac{\partial}{\partial X}\,\ln \langle \psi_{1,2}(W_t)\psi_{1,2}(X_t)\rangle_{\mathbb{T}_{p-t}}
\end{equation}
and therefore identify the stochastic differential equation for $W_t$:
\begin{equation}
  \label{eqn:drift}
  \diff W_t = \sqrt{\kappa}\,\diff B_t - \kappa\frac{\partial}{\partial X}\ln \langle \psi_{1,2}(W_t)\psi_{1,2}(X_t)\rangle_{\mathbb{T}_{p-t}},\quad X_t = g_t(x).
\end{equation}
The precise form of the two-point function appearing in this stochastic differential equation evidently depends on boundary conditions. In the sequel we shall concentrate on the scaling limit of LERWs.

\section{Application to LERWs}
\label{sec:lerw}

\subsection{The two-point function}
We now apply the preceding considerations in the case $\kappa=2$ in order to work out the passage probabilities. A conformal field theory in terms of simplectic fermions was suggested as suitable field-theoretic description of SLE$_2$ by Bauer, Bernard and Kyt\"ol\"a \cite{bauer:08}. The theory contains two basic free fermionic fields $\chi^+(z,\bar{z})$ and $\chi^-(z,\bar{z})$ for $z$ in some domain $\mathbb{D}$. Two-point function is given by $\langle \chi^\alpha(z,\bar{z})\chi^{\beta}(w,\bar{w})\rangle_\mathbb{D} = J^{\alpha\beta}G_{\mathbb{D}}(z,w)$ where $G_{\mathbb{D}}(z,w)$ denotes Green's function for the domain $\mathbb{D}$, and $J^{\alpha\beta}$ a $2\times 2$ antisymmetric matrix with $J^{++}=J^{--}=0$ and $J^{+-}=-J^{-+}=1$. The b.c.c. operator $\psi_{1,2}$ is related to the normal derivatives of these fermionic fields at the boundary of the domain $\mathbb{D}$. Without rephrasing their argument, we admit that the CFT correlation function $\langle\psi_{1,2}(0)\psi_{1,2}(x)\rangle_{\mathbb{T}_p}$ for this theory is equal to the \textit{excursion Poisson kernel} $H(x,p)$ in $\mathbb{T}_p$ for $\partial_1 \mathbb{T}_p$. It may be computed from Green's function for the Laplacian on $\mathbb{T}_p$ (see \ref{app:green}) and is given by
\begin{equation}
  \label{eqn:exkernel}
 \fl  H(x,p) = \frac{1}{2\pi \sin^2 (x/2)}-\frac{4}{\pi}\sum_{n=1}^\infty\frac{n\cos nx}{e^{2np}-1}-\frac{1}{p\pi}=-\frac{1}{\pi p}\left(p\, v'(x,p)+1\right).
\end{equation}
Using this together with \eqref{eqn:drift} leads to the stochastic differential equation \eqref{eqn:driftsde} for $\kappa=2$. This is consistent with Zhan's result on the scaling limit of planar LERWs on multiply-connected domains \cite{zhan:06_2}.
The identification $H(x,p)=\langle\psi_{1,2}(0)\psi_{1,2}(x)\rangle_{\mathbb{T}_p}$ implies that $H(x,p)$ must be solution of \eqref{eqn:bdpde} with $\mathcal{O}=\mathbf{1}$. Indeed, this can be seen from \eqref{eqn:defv}: since the Jacobi $\theta$-functions are particular solutions to Schr\"odinger's equation $4\pi i\,\partial\theta_1(z|\tau)/\partial\tau=\theta_1''(z|\tau)$, we conclude that $v(z,p)$ solves the (1+1)-dimensional \textit{Burgers equation}
\begin{equation}
  \label{eqn:burgers}
  \frac{\partial v(z,p)}{\partial p} = v(z,p)v'(z,p)+v''(z,p).
\end{equation}
Using this equation it is straightforward to derive
\begin{equation}
  \label{eqn:pdekernel}
 \fl \frac{\partial H(x,p)}{\partial p}= \left(\frac{\partial v(x,p)}{\partial x}+\frac{1}{p}\right)H(x,p)+v(x,p)H'(x,p)+H''(x,p).
\end{equation}
Comparison with \eqref{eqn:bdpde} for $\kappa=2$ and $h_{1,2}(\kappa=2)=1$ implies that the partition function is given by
\begin{equation}
  Z(p)= \frac{p}{\pi} \,\eta\left(\frac{ip}{\pi}\right)^2
  \label{eqn:partfunc}
\end{equation}  
up to a multiplicative constant. Here $\eta(\tau)$ denotes Dedekind's $\eta$-function $\eta(\tau)=e^{i\pi\tau/12}\prod_{n=1}^\infty (1-e^{2i\pi n\tau})$ \cite{abramowitz:70}.
Let us stress that \eqref{eqn:partfunc} is in accordance with Cardy's general result for the $O(n)$ partition function on annuli \cite{cardy:06}. Our case corresponds to $n=-2$.  In \ref{app:laplacian} we recall a relation between \eqref{eqn:partfunc} and the (regularised) determinant of the Laplacian on $\mathbb{T}_p$.

\subsection{Left-passage probability}
We are now in a position to write and solve the equations for the left-passage probability $\varpi(x,p)$. To this end, consider the conformal mapping $h_t(z)=g_t(z)-W_t$ where $g_t(z)$ is solution of \eqref{eqn:loewner}. Let us condition on a first portion $\gamma_t$ and evaluate the conditional probability $\mathsf{P}_{p,x}[\mathcal{L}\mathop{|}\mathcal{F}_t]$ that SLE$_2$ as defined above passes to the left ($\mathcal{F}_t$ is the filtration generated by $W_t$). We may use conformal invariance and the Markov property in order to write $\mathsf{P}_{p,x}[\mathcal{L}\mathop{|}\mathcal{F}_t]=\mathsf{P}_{p-t,h_t(x)}[\mathcal{L}]=\varpi(h_t(x),p-t)$. Hence
the process $\varpi(X_t=h_t(x),p-t)$ -- as a conditional probability -- must be a martingale for \eqref{eqn:driftsde}. Therefore, the application of It\^o's formula in combination with Loewner's equation \eqref{eqn:loewner} and \eqref{eqn:drift} yields the partial differential equation:
\begin{equation}
  \frac{\partial \varpi(x,p)}{\partial p}=\left[2(\log H(x,p))'+ v(x,p)\right]\varpi'(x,p)+\varpi''(x,p).
\end{equation}
If we write  $\lambda(x,p) = -p \pi H(x,p)\varpi(x,p)$ then it follows
\begin{equation}
  \label{eqn:fokkerplanck}
  \frac{\partial \lambda(x,p)}{\partial p}=\frac{\partial}{\partial x}\left(v(x,p)\lambda(x,p)+\frac{\partial \lambda(x,p)}{\partial x}\right).
\end{equation}
Remember that $v(x,p)$ is a particular solution of Burgers equation \eqref{eqn:burgers}. Hence we understand \eqref{eqn:fokkerplanck} in physical terms as evolution equation for a density advected in the Burgers velocity field $v(x,p)$.

In order to determine the initial conditions we must analyse $\lambda(x,p) = -p\pi\,H(x,p)\varpi(x,p)$ in the limit $p\to 0^+$. We expect that for $x\in [0,2\pi]$ $\lim_{p\to 0^+}\varpi(x,p) = \Theta(x-\pi)$, where $\Theta(x)$ denotes the Heaviside function which is $1$ for $x>0$, and $0$ for $x<0$. However, it shall be convenient to extend the function $\varpi(x,p)$ to $x\in \mathbb{R}$ by quasi-periodic continuation $\varpi(x+2\pi k,p) = \varpi(x,p)+k,\, k\in \mathbb{Z}, \, x\in [0,2\pi]$. It follows from \eqref{eqn:initcond} that $\omega(0,x) = k \in \mathbb{Z}$ for $(2k-1)\pi<x<(2k+1)\pi$. Moreover for all $x\notin 2\pi \mathbb{Z}$ we have $\lim_{p\to 0^+} p\,H(x,p)=0$ but it remains unclear how to handle the singularities at $x = 2\pi k,\,k\in \mathbb{Z}$. We take the naive limit of \eqref{eqn:exkernel} by writing $\lim_{p\to 0^+} (-p\pi)\,H(x,p) = 1+2\sum_{n\in \mathbb{Z}} \cos nx = 2\pi\sum_{n\in \mathbb{Z}}\delta(x-2\pi n)$ and justify our results \textit{a posteriori} by comparison to asymptotic limits. Thus, the initial condition for the density is given by
$\lambda(x,0) = 2\pi \sum_{n\in \mathbb{Z}} n\, \delta(x-2\pi n)$.

We now apply the technique which was used in \cite{hagendorf:08} where we  computed the winding number for SLE$_2$ on doubly-connected domains. For the primitive $\Omega(x,p)=\int_0^x\diff y\, \lambda(p,y)$ it follows
\begin{equation}
  \label{eqn:diffadv}
  \frac{\partial \Omega(x,p)}{\partial p}=v(x,p)\Omega'(x,p)+\Omega''(x,p)
\end{equation}
with initial condition $\Omega(0,x)\equiv\pi n(n+1) = \,\mathrm{const.}$ for $2n\pi < x < 2(n+1)\pi$. Because of the Burgers equation \eqref{eqn:burgers} we know that $v(x,p)$ itself satisfies \eqref{eqn:diffadv} and it is not difficult to see that the same holds for $p\, v(x,p)+ x$ \cite{zhan:04}. In order to find a general solution we set $\psi(x,p) = \theta_1(x/2\pi|ip/\pi)\Omega(x,p)$ (sometimes this is called a generalised Cole-Hopf transformation \cite{gorodtsov:99,ginanneschi:97,ginanneschi:98}). Then it is easy to show that $\partial \psi(x,p)/\partial p=\partial^2\psi(x,p)/\partial x^2$. Therefore $\psi(x,p)$ is solution of the simple diffusion equation which we may solve for any initial condition. In our case we have $\psi(x,0)=\theta_1(x/2\pi|0)\Omega(x,0)$. In fact, it is this simplification that leads to the nice explicit results in the case $\kappa=2$. Up to now, we are not aware of any similar transformation for $\kappa\neq 2$. After a little algebra, we recover the result for $\Omega(x,p)$:
\begin{eqnarray}
  \label{eqn:omega}
 \fl \Omega(x,p) &= \frac{1}{\theta_1(x/2\pi|ip/\pi)}\int_{-\infty}^\infty\frac{\diff y}{\sqrt{4\pi p}}\, e^{-(x-y)^2/4p}\theta_1(y/2\pi|0)\Omega(y,0)
  \nonumber\\
 \fl &=\frac{1}{\theta_1(x/2\pi|ip/\pi)}\sqrt{\frac{\pi}{p}}\sum_{n\in \mathbb{Z}}(-1)^n e^{-(x-\pi(2n+1))^2/4p} \Omega(\pi(2n+1),0)\nonumber\\
 \fl &= \frac{(x-\pi)(x+\pi)+2p}{4\pi}+\frac{p\,x}{2\pi}\,v(x,p)+\frac{p^2}{2\pi}\left(v'(x,p)+\frac{1}{2}v(x,p)^2\right).
\end{eqnarray}
In the second line, we use $\lim_{p\to 0^+}\theta_1(x/2\pi|ip/\pi)=2\pi\sum_{n\in \mathbb{Z}}(-1)^n\delta(x-\pi(2n+1))$.
Hence, upon derivation of $\Omega(x,p)$ with respect to $x$ and usage of Burgers equation \eqref{eqn:burgers} we find the density
\begin{equation}
  \lambda(x,p) = \frac{x}{2\pi}\left(p\,v'(x,p)+1\right)+\frac{p}{2\pi}\frac{\partial}{\partial p}\left(p\,v(x,p)\right).
\end{equation}
Eventually, this yields a closed form for the left-passage probability in terms of the Burgers velocity field:
\begin{equation}
  \label{eqn:proba}
  \varpi(x,p) = \frac{1}{2\pi}\left(x+\frac{\partial(p\, v(x,p))/\partial p}{v'(x,p)+1/p}\right).
\end{equation}
Since $v(x,p)$ is $2\pi$-periodic in $x$ we see that the solution indeed is quasi-periodic for any $p$. Let us check the asymptotic case as $p\to \infty$: we have $v(x,p)\sim \cot(x/2)$ up to exponentially small corrections of order $O(e^{-2p})$. If we neglect the latter we find
\begin{equation}
  \varpi(x,p)\mathop{\sim}_{p\gg 1}\frac{1}{2\pi}\left(x-\frac{\sin x}{1-2\sin^2(x/2)/p}\right)\mathop{\longrightarrow}_{p\to \infty}\frac{x-\sin x}{2\pi}.
\end{equation}
Thus on the right-hand side we see that the asymptotic limit yields the equivalent of Schramm's result \eqref{eqn:schramm} for $\kappa=2$ in our geometry. It is interesting that we find algebraic corrections in $1/p$ to the scaling limit and not only exponentially decaying terms. As $p\to 0^+$ we use the modular properties of $\theta_1(z|\tau)$ (see \ref{app:green})to show that for $x\in (0,2\pi)$ the probability $\varpi(x,p)$ is given by
\begin{equation}
  \label{eqn:initcond}
  \varpi (x,p)\mathop{\sim}_{p \to 0}
  \left\{
  \begin{array}{cc}
    e^{-2\pi(\pi-x)/p}, & 0 < x < \pi\\
    1-e^{-2\pi(x-\pi)/p}, & \pi < x < 2\pi\\
  \end{array}
  \right.
\end{equation}
up to corrections of $O(e^{-6\pi^2/p})$.
For $p=0$ we indeed have the Heaviside function $\Theta(x)$. The first corrections may be interpreted by a simple intuitive argument. On a lattice domain covering $\mathbb{T}_p$ consider a random walk from $0$ to $x$. For very narrow cylinders we expect that erasure of loops winding around the cylinder may be neglected to first order, and that we may find \eqref{eqn:initcond} from simple random walks in the limit $p\to 0^+$. Indeed, from a little computation of ``left-passage'' probability for random walks/Brownian motion to lowest order one retrieves \eqref{eqn:initcond}. 

\subsection{Conditioning LERWs and side arcs as target}
\label{sec:conditioning}

Stochastic Loewner evolutions with $\kappa=2$ have a special property pointed out by Bauer, Bernard and Kennedy \cite{bauer:08_2}. Consider a simply-connected planar domain $\mathbb{D}$ and fix a boundary point $x_0\in \partial \mathbb{D}$ as well as some side arc $S \subset \partial \mathbb{D}$. Stochastic Loewner evolutions in $\mathbb{D}$ from $x_0$ to $S$ are called \textit{dipolar} SLEs \cite{bauer:05}. The usual reference geometry is the strip $\mathbb{D}=\mathbb{S}_p$, $x_0=0$ and $S= \mathbb{R}+ip$. If we now condition dipolar SLE$_\kappa$ to hit some subinterval $I\subset S$ then we recover dipolar SLE$_\kappa$ in $\mathbb{D}$ from $x_0$ to $I$ if and only if $\kappa=2$. In particular, by shrinking $I$ to a point $x_\infty$, we obtain chordal SLE$_\kappa$ in $\mathbb{D}$ from $x_0$ to $x_\infty$ in $\mathbb{D}$ if and only if $\kappa=2$.

Here we aim at applying the method of conditioning in order to extend our previous result for SLE$_2$ on $\mathbb{T}_p$ from $0$ to $x$ to traces that are allowed to exit the cylinder at the side arc $S=[a,b]\subset \partial_1\mathbb{T}_p$, $0<a<b<2\pi$. Of course, thereby we tacitly assume that the results of \cite{bauer:08_2} also apply to multiply-connected domains. We shall denote the corresponding probability left-passage probability by $\Pi(a,b;\pi)$.

\paragraph{The limit $p\to +\infty$.} The cylinder $\mathbb{T}_{\infty}$ is conformally equivalent to the unit disc and may therefore be viewed as simply-connected domain. As previously mentioned, in this case the left-passage probability for curves from $0$ to $x$ may be computed by transport of Schramm's formula \eqref{eqn:schramm} to our geometry formula. It is given by $\varpi(x,\infty)=(x-\sin x)/2\pi$. If we fix some interval $S=[a,b]$ with $0<a<b<2\pi$ and consider dipolar SLE from $0$ to $S$, we may compute the left-passage probability as an integral from $a$ to $b$ of $\varpi(x,\infty)$ times the dipolar hitting density $\Lambda_{a,b}(x,\infty)$ at $x$. We obtain the latter from the excursion Poisson kernel \eqref{eqn:exkernel} as $p\to \infty$:
\begin{equation}
  \Lambda_{a,b}(x,\infty)= \frac{H(x,\infty)}{\int_a^b\diff x\,H(x,\infty)} = \frac{1}{2\sin^2(x/2)(\cot (a/2)-\cot (b/2))}.
\end{equation}
Hence we may write the left-passage probability as
\begin{equation}
  \label{eqn:piinfty}
  \Pi(a,b;\infty)=\int_a^b\hspace{-0.2cm}\diff x \,\varpi(x,\infty)\Lambda_{a,b}(x,\infty)=\frac{a\cot(a/2)-b\cot(b/2)}{\cot (a/2)-\cot(b/2)}.
\end{equation}
\paragraph{Application to the doubly-connected case.}
Let us now take a closer look at the doubly-connected case. We may follow the previous lines, however using the excursion Poisson kernel \eqref{eqn:exkernel} as well as the probability \eqref{eqn:proba}. This yields for the hitting density
\begin{equation}
  \Lambda_{a,b}(x,p) = \frac{H(x,p)}{\int_a^b\diff x\,H(x,p)}=\frac{-p\pi\,H(x,p)}{p(v(b,p)-v(a,p))+b-a}.
\end{equation}
This quantity allows for the evaluation of the left-passage probability $\Pi(p;a,b)$ for traces conditioned to exit on $[a,b]$:
\begin{equation}
  \Pi(a,b;p) = \int_a^b \hspace{-0.2cm}\diff x\, \varpi(x,p)\Lambda_{a,b}(x,p)=\frac{\Omega(b,p)-\Omega(a,p)}{p(v(b,p)-v(a,p))+b-a}
\end{equation}
with the functions $v(x,p)$ and $\Omega(x,p)$ are given in \eqref{eqn:defv} and \eqref{eqn:defomega} respectively. For large $p$ we find (up to corrections of $O(p\,e^{-2p})$
\begin{equation}
  \Pi(a,b;p) = \frac{a\cot(a/2)-b\cot(b/2)+(a^2-b^2)/2p}{\cot(a/2)-\cot(b/2)+(a-b)/p}
\end{equation}
and recover the result \eqref{eqn:piinfty} as $p\to+\infty$.

\section{Conclusion}
\label{sec:conclusion}
To conclude, in this article we have shown how to compute a left-passage probability for SLE$_2$ on doubly-connected domains. The feasibility of the computation, whose result constitutes a generalisation of Schramm's formula for $\kappa=2$, is intimately related to a particular case of a diffusion-advection problem in a Burgers velocity field as pointed out in our previous work \cite{hagendorf:08}. The application of conditioning for SLE$_2$ allowed for a generalisation of the left-passage probability to the case where the target interval is a side arc. It would be interesting in order to complete these
considerations and work out the probability that chordal SLE$_\kappa$ in $\mathbb{H}$ does not intersect a compact subset $\mathbb{K}\subset \mathbb{H}$. This problem was considered by R. Bauer in the special case $\kappa=8/3$ where the restriction property may be exploited \cite{rbauer:07}.

\section*{Acknowledgements}
The author wishes to thank Denis Bernard for very useful discussions and comments. Moreover, the author is grateful to Michel Bauer and Pierre Le Doussal for discussions, and Kay-J\"org Wiese for careful reading of the manuscript.

\appendix

\section{Excursion Poisson kernel}
\label{app:green}
Consider a planar (not necessarily simply-connected) domain $\mathbb{D} \subset \mathbb{C}$ and two boundary points $x_0,x_\infty\in \partial \mathbb{D}$. The excursion Poisson kernel is the partition function of Brownian paths on $\mathbb{D}$ with endpoints $x_0,x_\infty$. Notice that for multiply-connected domains, $x_0$ and $x_\infty$ may belong to different boundary components, what leads to different excursion Poisson kernels.

Here we recall the computation for $\mathbb{T}_p$ with $x_0=0$ and $x_\infty=x$ from Green's function. The propagator for two-dimensional Brownian motion started from $(x_0,y_0)$ in $\mathbb{T}_p$ is given by
\begin{equation}
  \fl \mathcal{P}(x,y,t|x_0,y_0;p){=} \sum_{n\in\mathbb{Z}}\frac{e^{-(x-x_0-2\pi n)^2/2t}}{\sqrt{2\pi t}}\frac{2}{p}\sum_{k=1}^\infty\sin\left(\frac{\pi k y}{p}\right)\sin{\left(\frac{\pi k y_0}{p}\right)}e^{-\pi^2k^2t/2p^2}
\end{equation}
We may find \textit{Green's function} $\mathcal{G}(x,y;x_0,y_0;p)$ for the Laplacian on $\mathbb{T}_p$ upon integration with respect to time \cite{lawler:05}:
\begin{eqnarray}
 \fl \mathcal{G}(x,y,x_0,y_0;p) = \int_0^\infty\hspace{-0.2cm}\rmd t\, \mathcal{P}(x,y,t|x_0,y_0;p)\\
 \fl=\frac{1}{p}\sqrt{\frac{2}{\pi}}\sum_{n\in \mathbb{Z}}\sum_{k=1}^\infty \sin\left(\frac{\pi k y}{p}\right)\sin\left(\frac{\pi k y_0}{p}\right)\int_0^\infty\hspace{-0.2cm}\rmd t\,\exp\left(-\frac{(x-x_0-2\pi n)^2}{2t}-\frac{\pi^2k^2t}{2p^2}\right).\nonumber
 \end{eqnarray}
Using the integral $\int_0^\infty \diff x\,[\exp(-ax-b/x)]/\sqrt{x}=\sqrt{\pi/a}\,\exp(-2\sqrt{ab})$ leads to
\begin{eqnarray}
 \fl \mathcal{G}(x,y,x_0,y_0;p)&=\frac{2}{\pi}\sum_{n\in \mathbb{Z}}\sum_{k=1}^\infty 
  \frac{\sin\left(\pi k y/p\right)\sin\left(\pi k y_0/ p\right)}{k}\exp\left(-\frac{\pi k |x-x_0-2\pi n|}{2p}\right)\nonumber\\
  & = -\frac{1}{\pi}\log\left|\frac{\theta_1((z-z_0)/2\pi|ip/\pi)}{\theta_1((z-\overline{z}_0)/2\pi|ip/\pi)}\right|-\frac{\mathrm{Im}\,z \,\mathrm{Im}\,z_0}{\pi p}
\end{eqnarray}
where $z=x+iy$, $z_0=x_0+i y_0$.
The excursion Poisson kernel is given by the normal derivative of Green's function at the boundary points $(x_0,0)$ and $(x,0)$ respectively. It only depends on $x-x_0$. From the preceding formula, we find
\begin{eqnarray}
 \fl H(x-x_0,p) &= \left.\frac{\partial^2\mathcal{G}(x,y,x_0,y_0;p)}{\partial y_0\partial y}\right|_{y=0,y_0=0}=\frac{\pi}{2p^2}\sum_{n\in \mathbb{Z}} \frac{1}{\sinh^2\left(\pi(x-x_0-2\pi n)/2p\right)}\nonumber\\
  \fl  &= \frac{\pi}{2p^2}\left(\frac{1}{\sinh^2(\pi (x-x_0)/2p)}+4\sum_{n=1}^\infty\frac{n \cosh (\pi n (x-x_0)/p)}{e^{2\pi^2n/p}-1}\right).
  \label{eqn:exkernel2}
\end{eqnarray}
Strictly speaking, the second line only converges if $0<|x-x_0|<2\pi$. In order to establish a relation with the series expansion given in \eqref{eqn:exkernel} recall the modular property $\theta_1(z|\tau)=\varepsilon \exp(-i\pi z^2/\tau)\theta_1(z/\tau|-1/\tau)/\sqrt{\tau}$ where $|\varepsilon|=1$. It allows to derive the formula
\begin{equation}
  v(x,p)=\frac{i\pi }{p}\,v\left(\frac{i \pi x}{p},\frac{\pi^2}{p}\right)-\frac{x}{p}.
\end{equation}
By virtue of this relation, it is easy to verify the equivalence of \eqref{eqn:exkernel} and \eqref{eqn:exkernel2}.

\section{Ward identities on doubly-connected domains}
\label{app:ward}
In this appendix we shall present a short derivation of the Ward identities for doubly-connected domains. We shall follow the strategy of \cite{bernard:88}, however see also \cite{eguchi:87} for conformal Ward identities on general Riemann surfaces.

\begin{figure}[h]
\centering
\begin{tikzpicture}[>= stealth]
  \draw (0.5,2) node {$\mathbb{H}$};
  \filldraw[lightgray] (0,0) rectangle (6,-0.2);
  \draw (0,0) -- (6,0);
  \draw[densely dashed,thick] (5,0) arc(0:180:2cm);
  \draw[densely dashed,thick] (4,0) arc(0:180:1cm);
  \draw[dashed,thick,gray] (2,0) arc(180:360:1cm);
  \draw[dashed,thick,gray] (1,0) arc(180:360:2cm);
  \filldraw[color=black] (4,0) circle (1pt);
  \filldraw[color=black] (5,0) circle (1pt);
  \filldraw[color=black] (3,0) circle (1pt);
  \filldraw[color=black] (3.4,1.3) circle (1pt);
  \filldraw[color=gray] (3.4,-1.3) circle (1pt);
  \draw (4.5,0.2) node {$\beta$}; 
  \draw (1.5,0.2) node {$\alpha$};
  \draw (3,1) node[rotate=60] {$=$};
  \draw (3,2) node[rotate=60] {$=$};

  \draw (3.8,-0.3) node {$q$}; 
  \draw (5.2,-0.25) node {$1$};
  \draw (2.8,-0.25) node {$0$};
  \draw (3.6,1.5) node {$w$};
  \draw[gray] (4.05,-1.1) node {$\bar{w}=w^\ast$};
  \draw[gray] (3,-1) node[rotate=60] {$=$};
  \draw[gray] (3,-2) node[rotate=60] {$=$}; 
  \draw[very thick] (1,0) -- (2,0);\draw[very thick] (4,0) -- (5,0);
\end{tikzpicture}
\caption{An doubly-connected domain is obtained from of a semi-annular domain on the upper half-plane $\mathbb{H}$ upon identification of $z=q e^{ix}$ and $z=e^{ix}$, $0\leq x \leq \pi$. Moreover, the picture illustrates the mirror image in the lower half-plane.}
\label{fig:semiannulus}
\end{figure}
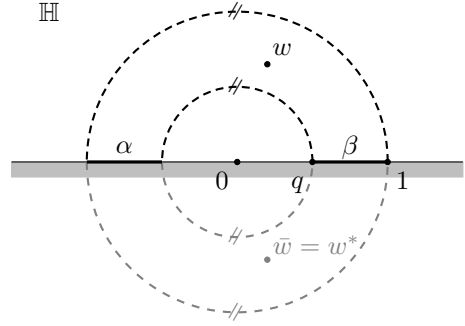

We start from boundary conformal field theory in the upper half-plane. In fact, for any $p$ we may map $\mathbb{T}_p$ to a semi-annular region in $\mathbb{H}$ with inner and outer radius identified as shown on figure \ref{fig:semiannulus}. Boundaries are given by the intervals $[-1,-q]$ and $[q,1]$ where $0<q<1$ is related to $p$ via $q=e^{-2\pi^2/p}$. On those two boundary components, different boundary conditions $\alpha$ and $\beta$ may be imposed. We use the operator formalism and radial quantisation in $\mathbb{H}$ \cite{cardy:84}.
Infinitesimal conformal transformations are implemented by appropriate insertions of the stress tensor $T(z)=\sum_{n\in \mathbb{Z}}L_nz^{-(n+2)}$ into CFT correlation functions. By Schwarz reflection $T(z)=\bar{T}(\bar{z})$ the theory in $\mathbb{H}$ is extended to the whole plane. However, this implies that we only work with a single copy of the Virasoro algebra. Its generators $L_n,\, n\in \mathbb{Z}$ satisfy the well-known commutation relation $[L_n,L_m]=(n-m)L_{n+m}+c n(n^2-1)\delta_{n+m,0}/12$ where $c$ denotes the central charge.
Dilatations are generated by the Virasoro operator $L_0$. Therefore we can write for the semi-annulus
\begin{equation}
  \langle\mathcal{O}\rangle = \frac{\tr \left(q^{L_0}\mathcal{O}\right)}{\tr \left(q^{L_0}\right)} = \frac{1}{Z}\tr \left(q^{L_0}\mathcal{O}\right)
  \label{eqn:cftcf}
\end{equation}
where $\mathcal{O}$ is any CFT observable. The trace operation implements the aforementioned identification of the semicircles with radii $q$ and $1$.

We sketch the derivation of the conformal Ward identity for doubly-connected domains using a single primary operator $\Phi(w,\bar{w})$ with $w\in \mathbb{H}$. The generalisation to an arbitrary product of different primary fields follows the same lines and is straightforward. Boundary conformal field theory in $\mathbb{H}$ prescribes that the dependence on the antiholomorphic variable $\bar{w}$ in the upper half-plane has to be interpreted as holomorphic dependence on $w^\ast$ in the lower half-plane. In fact, this is a reminiscent of Schwarz reflection. It may physically be understood as insertion of mirror images in the lower half-plane (see figure \ref{fig:semiannulus}). For a primary operator $\Phi(w,\bar{w})$ we have the commutation relation
\begin{equation}
\fl [L_n,\Phi(w,\bar{w})]
=\left[w^n\left(w\frac{\partial}{\partial w}+h(n+1)\right)+\bar{w}^n\left(\bar{w}\frac{\partial}{\partial \bar{w}}+h(n+1)\right)\right]\Phi(w,\bar{w}).
\label{eqn:commutator}
\end{equation}
Moreover, combination with the cyclicity of the trace operation in \eqref{eqn:cftcf}, the relation $[L_n,q^{L_0}]=(1-q^n)q^{L_0}L_n$ and the definition of the energy-momentum tensor $T(z)=\sum_{n\in \mathbb{Z}} L_n z^{-n-2}$ leads to
\begin{eqnarray}
  \tr (q^{L_0} T(z)\Phi(w,\bar{w}))-q\frac{\partial}{\partial q}\tr \left(q^{L_0}\Phi(w,\bar{w})\right)\nonumber\\
  \qquad = \left(G(z,w)\frac{\partial}{\partial w}+ h \frac{\partial G(z,w)}{\partial w}+(w\to \bar{w})\right)\tr q^{L_0}\Phi(w,\bar{w})
\end{eqnarray}
where the function $G(z,w)$ is defined by the series expansion
\begin{equation}
 \fl G(z,w)= \frac{w}{z^2}\sum_{n\neq 0}\frac{1}{1-q^n}\left(\frac{w}{z}\right)^n=\frac{w}{z^2}\left(\sum_{n=0}^\infty\frac{wq^n}{z-wq^n}+\sum_{n=-\infty}^{-1}\frac{z}{w-zq^n}\right).
\end{equation}
The second expansion may be considered as meromorphic continuation $G(z,w)$. As a function of $z$ we therefore have a meromorphic function with simple poles at $z=q^n w,\, n\in \mathbb{Z}$. Choosing the identity operator $\Phi(w,\bar{w})=\mathbf{1}$ with $h=0$, we find $\langle T(z)\rangle=
q\,\partial \ln Z(q)/\partial q$. Finally,
\begin{eqnarray}
  \fl\langle T(z) \Phi(w,\bar{w})\rangle{-} \langle T(z)\rangle\langle \Phi(w,\bar{w})\rangle=& \left( h \frac{\partial G(z,w)}{\partial w}+G(z,w)\frac{\partial}{\partial w}+(w\to \bar{w})\right)\langle\Phi(w,\bar{w})\rangle\nonumber\\
  &\hspace{2cm}
  +F(z)\frac{\partial}{\partial q}\langle \Phi(w,\bar{w})\rangle
\end{eqnarray}
where $F(z)=q/z^2$. For other geometries, consider a conformal mapping $f_q(z)$ from a doubly-connected domain $\mathbb{D}$ to the semi-annulus. Here we allow an explicit dependence of $f_q(z)$ on the modulus $q$. Then the functions $G$ and $F$ transform to some $\hat{G}$ and $\hat{F}$ respectively, and we have the transformation formulae
\begin{eqnarray}
 \hat{F}(z) = f_q'(z)^2 F(f_q(z))\\
  \hat{G}(z,w) = f_q'(z)^2 f_q'(w)^{-1}\left(G(f_q(z),f_q(w))-\frac{\partial f_q(w)}{\partial q}\, F(f_q(z))\right)\nonumber
\end{eqnarray}
Applying these rules to the conformal mapping $f_q(z)= -2\pi\log z/\log q$ we find $\hat{G}(z,w)= v(z-w,p)/2 + (\pi-z)/2p$. However, because of translation invariance of the correlation functions in $\mathbb{T}_p$ along the real axis we may drop the second term and work with $\hat{G}(z,w)= v(z-w,p)/2$. Moreover, we find $\hat{F}(z)= q(\log q)^2/4\pi^2$. Performing the change of variable $q\to p = -2\pi^2/\log q$ then leads to \eqref{eqn:ward}.

\section{Spectral determinant of the Laplacian}
\label{app:laplacian}
The partition function for the symplectic fermion theory for SLE$_2$ \cite{bauer:08} may be written in path-integral form
\begin{equation}
  Z = \int [\diff \chi^+][\diff\chi^-]\,\exp(-S[\chi^+,\chi^-])
\end{equation}
where the action is given by
\begin{equation}
  S[\chi^+,\chi^-] = \frac{1}{2\pi} \int_{\mathbb{D}} \diff^2z \, J_{\alpha\beta} \overline\partial\chi^\alpha(z,\bar{z})\partial\chi^\beta(z,\bar{z}),
\end{equation}
and $\mathbb{D}$ the planar domain the theory lives on.
Performing the path-integral yields $Z = \det (-\Delta)$. Hence the partition function is nothing but the (regularised) spectral determinant of the Laplacian. In this appendix we shall compute this quantity for $\mathbb{D}=\mathbb{T}_p$ by starting from the lattice and performing the scaling limit. Our method is a straightforward application of the results found in \cite{duplantier:88_2}.

To this end, we will consider a rectangular lattice approximation of the cylinder geometry. We introduce a lattice domain with integer coordinates $x\in \mathbb{Z}$ and $y=0,1,\dots,L$ and identify $x\equiv x+M$. Denote by $-\lambda_{\ell,m}$ the non-trivial eigenvalues of the discrete Laplacian $\Delta f(x,y)= f(x+1,y)+f(x-1,y)+f(x,y+1)+f(x,y-1)-4f(x,y)$ with vanishing boundary conditions on $y=0,L$. This choice implies $\lambda_{\ell,m}= 4- 2 \cos (\pi\ell/L)-2 \cos (2\pi m/M)$ with $\ell=1,2,\dots,L-1$ and $m=0,1,2,\dots,M-1$. Hence, the spectral determinant of $-\Delta$ is given by
\begin{eqnarray}
  \det (-\Delta)&=\prod_{\ell=1}^{L-1}\prod_{m=0}^{M-1}\lambda_{\ell,m}\nonumber\\
  &=\prod_{\ell=1}^{L-1}\prod_{m=0}^{M-1}\left(4- 2 \cos \frac{\pi \ell}{L}-2 \cos \frac{2\pi m}{M}\right)
  \label{eqn:detspec}
\end{eqnarray}
Our aim is to evaluate this expression as $M,L\to \infty$ in such a way that $p=2\pi L/M$ remains finite. To this end, we introduce a new variable $t_{\ell}$ with $\cosh t_\ell = 2-\cos \pi \ell/L$. More explicitly, we have
\begin{equation}
 t_{\ell} = \ln \left(2-\cos\left(\frac{\pi \ell}{L}\right)+\sqrt{\left(2-\cos\left(\frac{\pi \ell}{L}\right)\right)^2-1}\right)
\end{equation}
In particular, for fixed $\ell\ll L/\pi$ we may approximately write $t_\ell \approx \pi \ell/L$.

Evaluation of the product with respect to $m$ leads to
\begin{eqnarray}
  \prod_{m=0}^{M-1}\left(2\cosh t_\ell-2 \cos \frac{2\pi m}{M}\right) &= e^{M t_\ell}\prod_{m=0}^{M-1}|1-e^{-t_\ell+2\pi i m/M}|^2\nonumber\\
  &= e^{M t_\ell}(1-e^{-M t_\ell})^2.
\end{eqnarray}
The second line simply follows from the factorisation $z^M-1=\prod_{m=0}^{M-1}(z-e^{2\pi im/M})$. In order to proceed, we take the logarithm of \eqref{eqn:detspec} and combine it with the preceding result
\begin{equation}
  \ln \det (-\Delta) = \sum_{\ell=1}^{L-1}\left(M t_{\ell} + 2 \ln(1-e^{-M t_\ell})\right).
  \label{eqn:sum}
\end{equation}
Let us evaluate the first part of this sum by applying the well-known Euler-MacLaurin formula
\begin{eqnarray}
  \sum_{\ell=1}^{L-1} f(\ell) =& \int_{0}^L\diff x\, f(x)-\frac{f(0)+f(L)}{2}\nonumber\\&+\sum_{n=1}^m \frac{|B_{2n}|}{(2n)!}\left(f^{(2n-1)}(L)-f^{(2n-1)}(0)\right)+R_m.
\end{eqnarray}
Here $B_n$ is the $n^{\mathrm{th}}$ Bernoulli number. Their generating function is given by
$\sum_{n=0}^\infty B_n t^n/n!=t/(e^t-1)$. In particular, we have $B_2=1/6$. $R_m$ is some remainder that may be estimated \cite{abramowitz:70}. Applying this formula to $f(\ell)=M t_{\ell}$ yields
\begin{equation}
  \sum_{\ell=1}^{L-1}M t_\ell= \frac{4MLC}{\pi} -\frac{M}{2}\ln(3+2\sqrt{2})-\frac{\pi M}{12L}+\dots
\end{equation}
up to corrections decaying faster that $1/L$. Here $C = \sum_{n=0}^{\infty}(-1)^n/(2n+1)^2\approx 0.915966\dots$ is the Catalan constant.

The second part of the sum \eqref{eqn:sum} may be evaluated rather directly. In the limit $L\to \infty$ only the low $\ell$ values for $t_\ell$ play a decisive role. Hence we may approximate
\begin{equation}
  2\sum_{\ell=1}^{L-1} \ln(1-e^{-M t_\ell})\approx 2\sum_{\ell=1}^\infty\ln(1-e^{-\pi M\ell/L})
\end{equation}
up to some exponentially small corrections. Putting the two parts together yields -- after re-exponentiation --
\begin{equation}
  \fl\det(-\Delta) \approx \exp\left(\frac{4MLC}{\pi} -\frac{M}{2}\ln(3+2\sqrt{2})\right)\left(e^{-\pi M/24L}\prod_{\ell=1}^\infty (1-e^{-\pi M\ell/L})\right)^2.
\end{equation}
The regularised part of this result can be written in terms of Dedekind's $\eta$-function as $\left.\det(-\Delta)\right |_{\mathrm{reg.}} = \eta(i\pi/p)^2$. Using the modular property $\eta(\tau)=\eta(-1/\tau)/\sqrt{-i\tau}$ we find
\begin{equation}
  \left.\det(-\Delta)\right |_{\mathrm{reg.}} = \frac{p}{\pi}\,\eta\left(\frac{ip}{\pi}\right)^2
\end{equation}
as claimed in \eqref{eqn:partfunc}.


\section*{References}
\begin{thebibliography}{28}

\bibitem{schramm:00}
Oded Schramm,
\newblock {\em Scaling limits of loop-erased random walks and uniform spanning
  trees},
\newblock Israel J. Math. {\textbf{118}} (2000)   221--288.

\bibitem{lsw:04}
Gregory~F. Lawler, Oded Schramm  and Wendelin Werner,
\newblock {\em Conformal invariance of planar loop-erased random walks and
  uniform spanning trees},
\newblock Ann. Probab. {\textbf{32}} (2004)   939--995.

\bibitem{zhan:04}
Dapeng Zhan,
\newblock {\em {Stochastic Loewner evolutions in doubly connected domains}},
\newblock Prob. Theory and Relat. Fields {\textbf{129}} (2004)   340--380.

\bibitem{zhan:06}
Dapeng Zhan,
\newblock {\em {Some properties of annulus SLE}},
\newblock Electr. J. Prob. {\textbf{11}} (2006)   1069--1093.

\bibitem{zhan:06_2}
Dapeng Zhan,
\newblock {\em {The Scaling Limits of Planar LERW in Finitely Connected
  Domains}},
\newblock arXiv:math.PR/06100304 (2006).

\bibitem{ahlfors:73}
Lars~V. Ahlfors,
\newblock {\em {Conformal invariants: topic in geometric function theory.}},
\newblock McGraw-Hill, 1973.

\bibitem{lawler:05}
Gregory~F. Lawler,
\newblock {\em {Conformally Invariant Processes in the Plane}},
\newblock American Mathematical Society, 2005.

\bibitem{bauer:03}
Michel Bauer and Denis Bernard,
\newblock {\em {Conformal Field Theories of Stochastic Loewner Evolutions}},
\newblock Commun. Math. Phys. {\textbf{239}} (2003)   493--521.

\bibitem{bauer:06}
Michel Bauer and Denis Bernard,
\newblock {\em {2D growth processes: SLE and Loewner chains}},
\newblock Phys. Rep. {\textbf{432}} (2006)   115--221.

\bibitem{schramm:01}
Oded Schramm,
\newblock {\em A percolation formula},
\newblock Electr. Comm. Prob. {\textbf{6}} (2001)   115--120.

\bibitem{abramowitz:70}
Milton Abramowitz and Irene Stegun,
\newblock {\em {Handbook of Mathematical Functions}},
\newblock Dover Publications, 1970.

\bibitem{rbauer:08}
Robert~O. Bauer and Roland~M. Friedrich,
\newblock {\em {On Chordal and Bilateral SLE in multiply connected domains}},
\newblock Math. Z. {\textbf{258}} (2008)   241--265.

\bibitem{nehari:82}
Zeev Nehari,
\newblock {\em Conformal mapping},
\newblock Dover Publications, 1982.

\bibitem{bauer:04_2}
Michel Bauer and Denis Bernard,
\newblock {\em {SLE, CFT and zig-zag probabilities}},
\newblock in {\em Conformal Invariance and Random Spatial Processes}, NATO
  Advanced Study Institute, 2003.

\bibitem{difrancesco:97}
Philippe~Di Francesco, Pierre Mathieu  and David S\'en\'echal,
\newblock {\em {Conformal Field Theory}},
\newblock Springer, 1997.

\bibitem{cardy:84}
John Cardy,
\newblock {\em {Conformal invariance and surface critical behavior}},
\newblock Nucl. Phys. {\textbf{B240[FS12]}} (1984)   514--532.

\bibitem{eguchi:87}
Tohru Eguchi and Hirosi Ooguri,
\newblock {\em {Conformal and current algebras on a general Riemann surface}},
\newblock Nucl. Phys. {\textbf{B282}} (1987)   308--328.

\bibitem{bauer:08}
Michel Bauer, Denis Bernard  and Kalle Kyt{\"o}l{\"a},
\newblock {\em {LERW as an Example of Off-Critical SLEs}},
\newblock J. Stat. Phys. {\textbf{132}} (2008)   721--754.

\bibitem{cardy:06}
John Cardy,
\newblock {\em {The $O(n)$ Model on the Annulus}},
\newblock J. Stat. Phys. {\textbf{125}} (2006)   1--21.

\bibitem{hagendorf:08}
Christian Hagendorf and Pierre~Le Doussal,
\newblock {\em {SLE in doubly-connected domains and the winding of loop-erased
  random walks}},
\newblock J. Stat. Phys. (2008).

\bibitem{gorodtsov:99}
V.A. Gorodtsov,
\newblock {\em Convective heat conduction and diffusion in one-dimensional
  hydrodynamics},
\newblock JETP {\textbf{89}} (1999)   872.

\bibitem{ginanneschi:97}
Paolo~Muratore Ginanneschi,
\newblock {\em {Models of passive and reactive tracer motion: an application of
  Ito calculus}},
\newblock J. Phys. A: Math. Gen. {\textbf{30}} (1997)   L519--L523.

\bibitem{ginanneschi:98}
Paolo~Muratore Ginanneschi,
\newblock {\em {On the mass transport by a Burgers velocity field}},
\newblock Physica {\textbf{D115}} (1998)   341--352.

\bibitem{bauer:08_2}
Michel Bauer, Denis Bernard  and Tom Kennedy,
\newblock {Conditioning SLEs and loop-erased random walks},
\newblock arXiv:0806.2246 2008.

\bibitem{bauer:05}
Michel Bauer, Denis Bernard  and Jeremie Houdayer,
\newblock {\em {Dipolar stochastic Loewner evolutions}},
\newblock J. Stat. Mech. (2005)   P03001.

\bibitem{rbauer:07}
Robert~O. Bauer,
\newblock {\em {Restricting SLE$(8/3)$ to an annulus}},
\newblock Stochastic Processes and Their Applications {\textbf{117}} (2007)
  1165--1188.

\bibitem{bernard:88}
Denis Bernard,
\newblock {\em {On the Wess-Zumino-Witten models on the torus}},
\newblock Nucl. Phys. {\textbf{B303}} (1988)   77--93.

\bibitem{duplantier:88_2}
Bertrand Duplantier and Francois David,
\newblock {\em {Exact Partition Functions and Correlation Functions of Multiple
  Hamiltonian Walks on the Manhattan Lattice}},
\newblock J. Stat. Phys. {\textbf{51}} (1988)   327--434.

\end{thebibliography}
\end{document}